# Non-local common cause explanations for EPR

Matthias Egg and Michael Esfeld

Department of Philosophy, University of Lausanne

matthias.egg@unil.ch; michael-andreas.esfeld@unil.ch

(9 December 2013)

Abstract

The paper argues that a causal explanation of the correlated outcomes of EPR-type experiments is desirable and possible. It shows how Bohmian mechanics and the GRW mass density theory offer such an explanation in terms of a non-local common cause.

*Keywords*: Bell's theorem, Bohmian mechanics, EPR experiment, GRW mass density theory, local causality, non-local causes

*1.    Bell's theorem and the failure of local causality*

Bell's theorem (1964) (reprinted in Bell 2004, ch. 2) proves that any theory that complies with the experimentally confirmed predictions of quantum mechanics has to violate a principle of local causality. The idea behind this principle is that, in Bell's words, "the direct causes (and effects) of events are near by, and even the indirect causes (and effects) are no further away than permitted by the velocity of light" (Bell 2004, p. 239). This is one way of formulating the principle of local action that is implemented in classical field theories and that overcomes Newtonian action at a distance.

Consider the EPR experiment: two elementary quantum systems are prepared in an entangled state at the source of the experiment (such as two systems of spin 1/2 in the singlet state). Later, when they are far apart in space so that there is no interaction any more between them, Alice chooses the parameter to measure in her wing of the experiment and obtains an outcome, and Bob does the same in his wing of the experiment. Alice's setting of her apparatus is separated by a spacelike interval from Bob's setting of his apparatus. The following figure illustrates this situation:

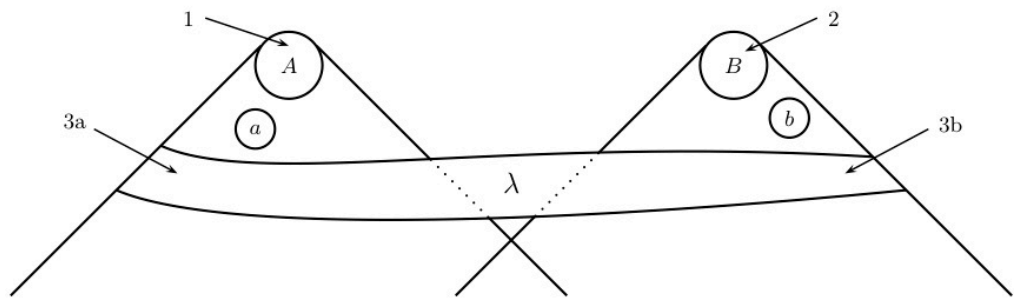

*Figure 1: The situation that Bell considers in the proof of his theorem. Figure taken from Seevinck (2010, appendix) with permission of the author.*

In this figure, *a* stands for Alice's measurement setting, *A* for Alice's outcome, *b* stands for Bob's measurement setting, *B* for Bob's outcome, and λ ranges over whatever in the past may influence the behaviour of the measured quantum systems according to the theory under consideration (which may be standard quantum mechanics, or a theory that admits additional, so-called hidden variables) (see Norsen 2009 and Seevinck and Uffink 2011 for precisions).

Bell's principle of local causality can then be formulated in the following manner:

$$P_{a,b}(A \mid B, \lambda) = P_a(A \mid \lambda) \quad (1)$$
$$P_{a,b}(B \mid A, \lambda) = P_b(B \mid \lambda)$$

That is to say: the probabilities for Alice's outcome depend only on her measurement setting and λ. Adding Bob's setting and outcome does not change the probabilities for Alice's outcome. The same goes for Bob. Bell's theorem then proves that quantum mechanics violates (1). Furthermore, any theory that reproduces the well-confirmed experimental predictions of quantum mechanics has to violate (1). This conclusion applies not only to quantum mechanics, but also to quantum field theory (see Bell 2004, ch. 24, and Hofer-Szabó and Vescernyés 2013 as well as Lazarovici forthcoming for the current discussion). One can therefore say that Bell's theorem puts a constraint on any – present or future – physical theory that is to match the experimentally confirmed predictions of quantum mechanics.

Note that the proof of Bell's theorem requires not only the principle of local causality, but also that the measurement settings *a* and *b* are independent of λ. Failure of such independence can arise in two different ways: either the measurement settings exert some influence on λ, or λ somehow influences the measurement settings. It is obvious from Figure 1 that the first option involves influences travelling backwards in time (see Price 1996, ch. 8 and 9, as well as the papers in *Studies in History and Philosophy of Modern Physics* 38 (2008), pp. 705-784). The second option contradicts the presupposition that the measurement settings can be freely chosen (see Bell et al. 1985).

Since the presupposition of an independence between the measurement settings and the prior state of the measured system is not specific for Bell's theorem, but applies to any experimental evidence, it is reasonable to conclude that this theorem establishes a violation of the principle of local causality (see Maudlin 2011, chs. 1-6, for a detailed assessment). However, it would be premature to infer from the failure of local causality that no causal explanation of the correlated measurement outcomes in the EPR experiment is possible. Suárez (2007) reviews the most influential arguments for this claim and shows that they are all inconclusive. To be more precise, one cannot expect an explanation of why nature is non-local (in the same way as one cannot expect an explanation of why local causality holds in classical field theories). But the fact that nature is non-local does not prevent one from searching for a causal explanation of the correlated outcomes of an EPR-type experiment. This fact only shows that it would be futile to search for an explanation in terms of causes conforming to the principle of local action.

In the next two sections, we will first ask what kind of causal explanation we should seek for the correlated outcomes of the EPR experiment (section 2) and then consider a general scheme for a causal law (section 3). Against this background, the heart of the paper is to argue that we have at least two elaborate proposals at our disposal that offer an explanation of the EPR correlations in terms of a non-local common cause, namely Bohmian mechanics (section 4) and the GRW mass density theory (section 5).

## 2.    *Direct or common cause?*

There are two principled ways of causally explaining correlations between two distinct events: one can either postulate a direct causal influence of one event on the other or one can suppose that a common cause of the two events accounts for the correlation. In everyday contexts, this distinction is crucial, because it has consequences for the kinds of manipulations that we can perform. If there is a direct causal link from event *A* to event *B*, then we can (in principle) bring about changes in *B* by intervening on *A*, whereas this is not possible if *A* and *B* are only connected via a common cause. In the context of quantum mechanics, however, the distinction between a direct cause (DC) and a common cause (CC) is much more elusive, since we cannot control the outcomes of quantum measurements in the right way to perform the intervention on *A* that is necessary to distinguish between these two causal structures.

Nevertheless, there is a metaphysical interest in separating DC from CC explanations for the EPR correlations, because these two types of explanation involve very different kinds of non-locality. (That there has to be *some* kind of non-locality is a consequence of Bell's theorem, as discussed in the previous section.) A CC explanation of the EPR correlations is committed to some kind of holism or non-separability in the sense that the common cause, when bringing about the outcome in one wing of the experiment, has to take into account what happens in the other wing. This holistic character of the common cause, which somehow spans both wings of the experiment, may seem unfamiliar. By contrast, the non-locality involved in the DC model seems metaphysically much more palatable: such a model can be completely separable and no spatially extended causes are needed; the only unusual aspect is that some causal influences travel faster than light. However, it is important to note that the DC model only has this advantage as long as the speed of the supposed causal influence is finite. Consider what van Fraassen points out in this context:

> To speak of instantaneous travel from *X* to *Y* is a mixed or incoherent metaphor, for the entity in question is implied to be simultaneously at *X* and at *Y* – in which case there is no need for travel, for it is at its destination already. … one should say instead that the entity has two (or more) coexisting parts, that it is spatially extended. (van Fraassen 1991, p. 351)

Hence, if the speed of the supposed causal influence is infinite, the DC model no longer displays separability, because the causal influence then is at two places at the same time. We therefore take the finite speed condition to be an essential part of any DC model. Due to their non-locality, both the DC and the CC model are in tension with special relativity, by attributing objective reality to some frame dependent notions (the speed of the causal influence in the DC model, and the commitment to objective simultaneity in the CC model). Both models are thus implicitly committed to a preferred foliation of space-time.

One might think that by opting for the DC model, one incurs less ontological costs than by accepting the holism of the CC model. Nevertheless, there is at present no serious proposal for explaining the EPR correlations along the lines of the DC model. Nor does there seem to be any strong motivation for developing such proposals, as recent research has brought forth a number of empirical and theoretical results which render the DC model increasingly unattractive:

 (a) Large-distance Bell tests put a substantial lower bound on the speed of the causal influence *v* postulated by the DC model. For an earth-based reference frame, Zbinden et al. (2001) found $v > 2/3 \cdot 10^7 c$, where *c* denotes the speed of light. More recent experiments generalize these results to other reference frames, and the lower bound on *v* is likely to increase further as better measurement techniques become available (see Salart et al. 2008 as well as Cocciaro, Faetti and Fronzoni 2011 and 2013). Due to this ongoing experimental research, the finite speed assumption becomes more and more doubtful.

(b) Some natural ways of implementing the DC model by choosing a suitable reference frame are even completely ruled out by experimental data. If one entertains the idea that the preferred reference frame is determined by the experimental set-up, then one has to consider the possibility that the two measuring devices in an EPR experiment (or at least their relevant parts) are in relative motion, thus picking out two *different* reference frames. It then becomes possible to realize a so-called *before-before* configuration, in which each of the two measurements, in the rest frame of the corresponding measuring device, is considered to happen before the other one. Since this implies that neither of the two events can influence the other one, the DC model predicts the disappearance of EPR correlations in such a set-up. Zbinden et al. (2001) and Stefanov et al. (2002) have empirically falsified this prediction.

(c) As long as only two-particle systems are considered, a DC model can always reproduce the quantum mechanical predictions by a suitable choice of the preferred reference frame and a sufficiently high value of *v*. However, as Bancal et al. (2012) have shown, this is no longer the case for systems of four entangled particles. In particular, any DC model then predicts a violation of the so-called no-signalling conditions.

(d) Part of the attractiveness of the DC model stems from its simplicity: it seems that one only needs to postulate a causal influence from one measurement event to the other one. However, Näger (2013) has shown that this is not the case. In order to violate Bell's inequalities, a causal model must at least postulate a causal influence from one of the measurement settings (e.g., the event we denoted by *a* in section 1) to the distant outcome (*B*), which does not go through the local outcome (*A*). This does not rule out all DC models, but it shows that they cannot be as simple as is usually thought.

We conclude from these results that a plausible causal explanation of the EPR correlations is not to be sought along the lines of the DC model. Rather, the explanation needs to invoke a (non-local) common cause (see Hofer-Szabó, Rédei and Szabó 2013, ch. 9, for a detailed investigation of the conditions under which common cause explanations for EPR correlations are possible). Before considering concrete proposals in that sense, we introduce a general conceptual scheme for causal explanations in the next section.

## 3.    A general scheme for causal explanations

Consider Newtonian mechanics. This theory seeks to explain the temporal development of the velocity of particles that are localized at points in physical space. The theory accounts for change of velocity in terms of forces acting on the particles, and these forces are traced back to properties of the particles. Thus, in virtue of instantiating the property of gravitational mass, particles exert a force of attraction upon each other. If one asks why the particles change their state of motion, the answer is that they do so because they possess gravitational mass. Gravitational mass can therefore be conceived as a causal property or a disposition whose effect or manifestation consists in the force that a particle exerts upon other particles in virtue of possessing gravitational mass.

Regarding gravitational mass as a dispositional property that figures in a causal explanation of the mutual attraction of massive particles does not imply the commitment to a particular metaphysics of properties or a particular stance in the metaphysics of laws of nature. One can notably leave open whether it is essential for the property of gravitational mass that it exerts the causal role of attracting other particles (dispositional essentialism – see notably Bird 2007) or whether it is a contingent feature of this property that it plays this causal role in the actual world (Humeanism – see notably Lewis 2009). Furthermore, one can leave open whether the property of gravitational mass grounds Newton's law of gravitation (or whatever may be the correct law of gravitation) in the sense that the law supervenes on this property – so that in every possible world in which the property of gravitational mass is instantiated, Newton's law of gravitation holds. In brief, endorsing the mentioned causal explanation requires only to acknowledge that the property of gravitational mass exerts a certain causal role in the actual world.

Referring to the property of gravitational mass instantiated by the particles provides for a causal explanation of the acceleration of the particles independently of whether or not a medium is indicated by means of which the influence that particles exert on each other's state of motion is transmitted and independently of whether or not time passes between the presence of gravitational mass (the cause) and the acceleration of the particles (the effect). Thus, Blondeau and Ghins (2012) argue that the "general form of a causal law is an equation that exhibits the following mathematical form:

$$E = \partial x / \partial t = C_1 + \ldots C_n \qquad [(2)]$$

$E$ refers to the effect, whereas the causes $C_i$ can, but need not, be functions of time. The above general form reads: $C_1$, $C_2$, … are the causes of the infinitesimal variation of the property $x$ of a system, i.e. of the effect $E$" (p. 384). The decisive point is that any law fitting into this form is asymmetric in that what appears on the right side induces a certain temporal development of the quantity on the left side, but not *vice versa*, without any time passing between the presence of the causes $C_1 \ldots C_n$ and the effect $E$, that is, the manner in which $x$ develops in time. Thus, on Newton's law of gravitation, the presence of gravitational mass induces a change in the velocity of the particles without any time passing between the presence of mass and the acceleration of the particles. Furthermore, for there being a causal law, providing for causal explanations, it is sufficient to indicate properties $C_1 \ldots C_n$ that have the effect of determining the temporal development of the value of the quantity $x$, but it is not necessary to indicate a medium through which they do so. These two points – no time passing between the cause and its effect, no medium of transmission necessary – will be crucial in developing an explanation of the EPR correlations in terms of a non-local common cause.

The proposal set out by Blondeau and Ghins (2012) shows that causal explanations can be given all over physics. If such explanations are cast in terms of (2), there is no point in claiming that the search for them is rooted in methodological and metaphysical preconceptions that modern physics has outgrown. In particular, the availability of what is known as structural explanations by no means makes the search for causal explanations superfluous (as claimed, e.g., by Dorato and Felline 2011). The interest in causal explanations is to obtain an answer to the question of why a certain variable develops in a certain manner in time – or, in other words, why a certain event occurs at a certain place and time. Thus, on the account of Blondeau and Ghins (2012), there even is a causal explanation of inertial motion, namely in terms of the initial velocity (p. 396) – although inertial motion is often cited as a prime example of a phenomenon no longer calling for a causal explanation (see e.g. Dorato and Felline 2011, p. 170). By the same token, even if one were to accept the claim that quantum mechanics includes a structural account of EPR-type correlations, the search for a causal explanation of these correlations would still be well motivated.

Before going into such an explanation, let us briefly consider the other force treated in classical physics, namely the electromagnetic force. The scheme is the same as in the case of gravitation: in virtue of instantiating the property of positive or negative charge, charged particles exert a force of attraction or repulsion upon each other. If one asks why the particles change their state of motion, the answer is that they do so because they possess positive or negative charge. Charge can therefore be conceived as a causal property or a disposition whose effect or manifestation consists in the change of the state of motion that a particle induces in other particles in virtue of being charged.

By contrast to Newtonian gravitation, the force that a particle exerts upon other particles in virtue of being charged is retarded, and it is transmitted through a medium, the electromagnetic field. Thus, charged particles generate a field – the electromagnetic field – that transmits the electromagnetic force, so that the effect of a charged particle on the velocity of other charged particles is retarded, the velocity of light being a constant that constitutes the upper limit velocity for the propagation of that effect. The classical theory of electromagnetism is therefore generally seen as setting the paradigm for causal explanations in terms of local action, that is, retarded action transmitted by a medium. Against this background, Einstein is considered as accomplishing this paradigm in the special theory of relativity and as developing a theory of gravitation in terms of local action in the general theory of relativity, by identifying the gravitational field with the metrical field of space-time. Nonetheless, even if one endorses this point of view, it is by no means necessary for a causal explanation to fit into the paradigm of local action, as the above mentioned general scheme of a causal law and the example of Newton's theory of gravitation show. In other words, it is an empirical question whether or not causal explanations follow the paradigm of local action. Bell's theorem proves that they cannot do so in quantum physics.

*4.     Non-local common cause I: Bohmian mechanics*

When turning to quantum physics, one has to be aware of the fact that what the textbooks provide is a formalism in terms of the temporal development of a wave-function of quantum systems, which enables the calculation of probabilities for measurement outcomes by means of defining operators or observables and which tells us how these probabilities develop in time. But the textbooks do not spell out what in the physical world the wave-function and its temporal development represent. The formalism of a wave-function and its temporal development cannot even take into account the fact that there are measurement outcomes, unless something is added to that formalism. If one takes for granted that there are measurement outcomes in physical space, there then are two principled possibilities to add something to the core textbook formalism of quantum mechanics: one can either recognize the Schrödinger equation as the law for the temporal development of the wave-function and provide an additional law that links the wave-function up with the distribution of matter in space-time and its temporal development, or one can change the Schrödinger equation so that a law is achieved that can accommodate the fact of there being measurement outcomes. The paradigmatic examples are Bohm's quantum theory for the former approach and the theory of Ghirardi, Rimini and Weber (GRW) for the latter approach. We will now show how each of these theories provides for an explanation of the EPR correlations in terms of a non-local common cause.

Bohm's theory proposes an ontology of particles located in physical space and a law of motion that describes the temporal development of the position of the particles, the so-called guiding equation. While the actual particle configuration is not taken into account by the quantum mechanical wave-function (that is why the position of the particles in Bohm's theory is often referred to as a hidden variable), the wave-function and its temporal development according to the Schrödinger equation, if put into the Bohmian guiding equation and applied to the actual particle configuration, enable to define a velocity field along which the particles move. Bohm himself presented this theory as a causal approach to quantum physics (see notably Bohm 1952, Bohm and Hiley 1993, as well as Holland 1993). He considered his theory as causal, because he showed that one can understand quantum mechanics within the paradigm that Newtonian mechanics set for modern physics, namely by adding a further force that accounts for the situations in which the trajectories of the particles are not correctly described by classical mechanics. That new force is known as the quantum potential.

However, the commitment to the quantum potential is widely seen as an *ad hoc* move to cast quantum mechanics in the framework of classical mechanics, because the quantum potential is quite unlike a classical force. The main objections are the following three:

(a) Unlike a classical force, the quantum potential does not satisfy Newton's third law: there is no reaction from the particles that corresponds to the action of this force on them.

(b) Unlike the classical forces of gravitation and electromagnetism, the quantum potential cannot be traced back to a property that each of the particles instantiates (like mass and charge). It cannot be conceived as a field either, for it does not have a value at points in space-time. The wave-function, which is supposed to represent the quantum potential, does not permit to assign values to points of physical space-time; if it is a field, it can be a field only on configuration space, that is, the mathematical space each point of which corresponds to a possible configuration of the particles in physical space. However, it is unintelligible how a force field on configuration space could move particles in physical space.

(c) Unlike the whole of classical physics, the core of Bohm's theory does not consist in a second order equation fulfilling the scheme set by Newton's second law, namely to employ forces in order to explain the temporal development of the velocity of particles. Instead, it is a first order equation, which can be written down in the following manner:

$$\frac{dQ}{dt} = \mu \Im \frac{\nabla \Psi_t(Q)}{\Psi_t(Q)} \qquad (3)$$

In this law, the quantum mechanical wave-function $\Psi_t$ has the job to determine the velocity of the particles at a time $t$, given their position $Q$ at $t$, with $\Im$ denoting the imaginary part and $\mu$ being an appropriate dimension factor.

Let us therefore leave the scheme of causal explanations in terms of forces that can be traced back to properties of particles behind and ask how today's dominant version of Bohm's theory can explain the measurement outcomes of quantum physics. This version, known as Bohmian mechanics, is committed only to particles localized in physical space and a law of motion (the guiding equation) in which the wave-function is employed in order to describe how the positions of the particles develop in time (see the papers collected in Dürr, Goldstein and Zanghì 2013). The wave-function figuring in equation (3) is the universal wave-function of all the particles in the universe. Since this wave-function is entangled, the law of motion of Bohmian mechanics is non-local: the velocity of any particle at a time $t$ depends, via the wave-function, on the position of all the other particles at $t$. But this dependency does not mean that there is an interaction among the particles (which would then be an instantaneous action at a distance). It only means that the temporal development of any particle is correlated with the temporal development of all the other particles through the wave-function (although, due to the decoherence of the universal wave-function, that dependence can in many cases be neglected). Dürr, Goldstein and Zanghì therefore regard the universal wave-function as nomological, its ontological status being limited to the role that it performs in the guiding equation (2013, ch. 12). Of course, simply attributing a nomological status to the wave-function leaves open all the questions about how laws of nature are grounded in physical reality. But this move makes it possible to adopt towards the guiding equation and whatever figures in it the same attitude as outlined in the preceding section. In short, one can consider the guiding equation as a causal law.

Let us come back to the general form of a causal law (2). If the ontology is one of particles and if the law is a first order equation, then its left side is about the temporal development of the position of the particles ($dQ/dt$) and its right side indicates the *beables*, to use Bell's terminology, that determine (in a deterministic or a probabilistic manner) their velocity. That is to say, the quantum mechanical wave-function represents a beable that fixes the velocity of the particles. That beable is also known as the quantum state. Apart from the difference between a second order and a first order theory, the only difference from classical mechanics is that this beable is a non-local one, instead of a local beable instantiated by each particle (such as its mass and its charge), and that this beable itself develops in time. As mentioned in the quotation from Blondeau and Ghins (2012) in the preceding section, the general form of a causal law (2) admits that the factors appearing on the right side also develop in time – as does the wave-function (unless it should turn out that the universal wave-function is stationary, as demanded by the Wheeler-DeWitt equation in quantum gravity).

To give a concrete meaning to this notion of a non-local beable in the framework introduced in sections 2 and 3, one can despite its non-local character maintain that it is a causal property, namely a holistic and dispositional property of the configuration of all the particles, with the manner in which the position of the particles develops in time being the manifestation of this disposition (see Belot 2012, pp. 77-80, and Esfeld et al. 2013 for a detailed exposition of the view of the wave-function representing a dispositional property). In the following, we shall take up this conception of the non-local beable represented by the wave-function being a holistic property of all the particles in order to provide a concrete meaning to our proposal of a non-local common cause. But this proposal by no means depends on spelling out that non-local beable in terms of a (holistic and dispositional) property of the configuration of matter in physical space. For the argument of this paper, one can also simply leave it at saying that the wave-function represents a non-local beable. In any case, this non-local beable is something over and above the local beables consisting in the position of matter in physical space. In other words, the property of position of the particles does not determine the holistic property of the particle configuration that is represented by the wave-function and that fixes the temporal development of the particles.

Hence, in brief, non-locality notwithstanding, Bohmian mechanics offers a causal explanation of the behaviour of the particles in terms of a non-local beable that can be conceived as a holistic, causal property instantiated by the particle configuration and represented by the wave-function. This explanation is less intuitive than an explanation in terms of forces acting upon the particles. However, the point at issue is to come up with a conceptually clear and coherent causal explanation, instead of satisfying anthropomorphic intuitions about agent-like entities acting upon each other. The latter ones have been criticised with good reason by Russell (1912) in his famous denunciation of the notion of causality. Despite Russell's criticism, there is a clear notion of a causal law, a causal property and a causal explanation in physics, but it is simply one in terms of the variables that determine the temporal development of physical quantities (such as the position, or the velocity of particles).

Let us now see how this causal explanation works in the case of the EPR experiment. Bohmian mechanics is a deterministic theory. That is to say, measurement outcomes are determined before they occur. What determines the correlated outcomes of an EPR experiment is the initial particle configuration – that is, the exact initial position of each of the two particles at the source of the experiment – plus the setting of the parameters that are measured in each wing of the experiment. These elements determine the measurement outcomes via, strictly speaking, the universal wave-function, whereby it is in this case sufficient to consider the effective wave-function of the particle pair (which depends on the settings). In other words, the two quantum particles and the particles constituting the settings of the parameters in both wings of the experiment instantiate a non-local beable that can be conceived as a causal property or a disposition, which is represented by the wave-function and whose effect or manifestation are the correlated measurement outcomes.

The common cause of the measurement outcomes is non-local, because it depends on the setting of the parameters in both wings of the experiment. More precisely, if one enquires into the cause of the measurement outcome in one wing and if this outcome occurs later than the outcome in the other wing on the absolute time parameter that Bohmian mechanics presupposes, that cause depends on both settings, that is, also the setting in the *other* wing. Bohmian mechanics therefore violates the condition known as parameter independence, as any deterministic theory has to do (see Jarrett 1984, Shimony 1993, pp. 144-149, and see also Norsen 2009 and Näger 2013 for a critique of Jarrett's analysis).

Since this common cause explanation postulates a causal dependence of the outcome in one wing on the setting in the other wing, one might ask whether it does not allow for superluminal signalling. In Bohmian mechanics, such signalling is usually taken to be ruled out by the fact that it is impossible to obtain precise knowledge of the initial conditions, that is, the exact initial position of each of the two particles at the source of the experiment. However, given that we are here dealing with a causal dependence between a controllable factor (the setting) and an observable one (the outcome), it is not immediately clear why our lack of knowledge about a *further* factor (the exact initial configuration) should prevent us from exploiting this dependence to send signals. Indeed, Bohmian mechanics only avoids the possibility of superluminal signalling by relying on what is known as the *quantum equilibrium hypothesis*, which connects the initial distribution of the particle positions with the wave-function (see Dürr, Goldstein and Zanghì 2013, ch. 2). If the equilibrium hypothesis did not hold, one could send a signal by manipulating the measurement settings in one wing, thereby influencing the effective wave-function of the two particles, which in turn would influence the outcome in the other wing. But due to quantum equilibrium, the spatial distribution of the particles is such that the change in the effective wave-function does not result in a change of the measurement outcome statistics in the distant wing. Wood and Spekkens (2012, sec. IV C) take this to be a case of fine-tuning. If, however, the quantum equilibrium hypothesis can be justified in terms of the statistical behaviour arising from a typical initial configuration, as Dürr, Goldstein and Zanghì (2013, ch. 2) argue, there is nothing problematic about this fine-tuning.

Bohmian mechanics is thus not committed to superluminal causation in an operational sense, but it is so committed in a metaphysical sense: given any initial particle configuration, the theory supports counterfactual claims of the type: "If Alice had chosen a different setting, Bob would have obtained a different outcome". This might sound like a kind of action at a distance that should be understood in terms of a DC model (see section 2), rather than as the manifestation of a common cause. We nevertheless consider the CC point of view more appropriate, for two reasons: First, as argued in section 2, the DC model includes a commitment to the finite speed of the causal influence, in contrast to the Bohmian explanation developed here. Second, it is not correct to say that there is a *direct* causal influence from Alice's setting to Bob's outcome. Alice's setting influences Bob's outcome only *via a further causal factor* (wich in turn influences *both* outcomes), namely the non-local beable (or holistic property) represented by the effective wave-function of the two particles. This factor, together with the (spatial) particle configuration, constitutes the common cause of the measurement outcomes.

Since the common cause in this explanation depends on the settings of the measurement parameters, it is obvious that it cannot be identified with the variable $\lambda$ in figure 1, which is independent of the measurement settings. Rather, $\lambda$ here includes only the initial quantum state (represented by the initial wave-function) and the initial configuration of the two particles, whereas the common cause is a later (quantum plus configuration) state of the two particles, which depends on the previously chosen measurement settings. Our explanation therefore violates what San Pedro (2012) calls "measurement independence", but is often (rather misleadingly, as he argues) called "no-conspiracy condition". We agree with San Pedro that this violation does not imply any kind of conspiracy, and our explanation has a similar structure to the common cause model he proposes. We disagree, however, about the sharp contrast he draws between his model and Bohmian mechanics, based on the claim that the latter, unlike the former, satisfies measurement independence (p. 154). This claim seems to rest on the (mistaken) identification of the common cause with the variable $\lambda$. Our analysis shows that if one correctly identifies the common cause, Bohmian mechanics violates measurement independence just as San Pedro's causal model does.

The difference between $\lambda$ and the common cause in the explanation given above is also important if one tries to connect our discussion with some other statistical conditions familiar from the literature on Bell's theorem. For example, if one were to replace $\lambda$ in Bell's locality condition (1) by the common cause variable, the deterministic character of our model would straightforwardly imply the fulfillment of this condition (both sides of each equation becoming either 0 or 1). By the same token, our above statement that Bohmian mechanics violates parameter independence would then become false. However, this should not mislead anyone into calling our causal model "local": the appearance of locality simply arises because the (non-local) dependence on the measurement settings has been absorbed into the common cause variable. To reiterate this point, using the terminology of Hofer-Szabó, Rédei and Szabó (2013, ch. 9): the fact that our model satisfies "hidden locality" has no metaphysical significance, since it simultaneously violates "weak no-conspiracy". Finally, let us note in this context that we share the widely held intuition that a common cause should function as a *screener-off* for the correlation it is supposed to explain. The common cause in our explanation does this, again, by virtue of its deterministic character.

If the EPR experiment is done with two particles of spin 1/2 in the singlet state – so that the measurement outcomes consist in correlated values of spin in certain directions –, nothing in this explanation changes. According to Bohmian mechanics, spin is not a local beable over and above position, but a manner in which particles behave in certain experimental contexts (see Bell 2004, ch. 4, and Norsen 2013). The only local property that the particles have is their position, and all measurement outcomes consist in a physical entity having a certain position in space at a certain time. Thus, the correlated outcomes of an EPR experiment on two particles of spin 1/2 in the singlet state are in Bohmian mechanics also accounted for by the mentioned common cause, namely a non-local beable (a holistic property) instantiated by the two quantum particles and the particles constituting the two settings, which determines the measurement outcomes in determining how the particles move.

5. *Non-local common cause II: the GRW mass density theory*

Let us now turn to the other type of approach to conceive the quantum formalism as representing the distribution of matter in physical space, namely to change the Schrödinger equation so that a law is achieved that can accommodate the fact of there being measurement outcomes in physical space, as exemplified by the formalism proposed by Ghirardi, Rimini and Weber (1986) (GRW). An equation that is able to describe measurement outcomes does of course in itself not reveal how the distribution of matter in physical space constitutes measurement outcomes. In other words, it does not tell us what the matter in space and time is, which is represented by the GRW formalism. Ghirardi answers this question by proposing an ontology of a continuous distribution of matter in space – a mass density field – that develops in time according to the GRW law (see Ghirardi, Grassi and Benatti 1995, and see Monton 2004 for a philosophical discussion). More precisely, the wave-function in configuration space and its temporal development according to the GRW equation represent at any time the density of matter (mass) in physical space. The spontaneous localization of the wave-function in configuration space represents a spontaneous contraction of the mass density in physical space, thus accounting for measurement outcomes and well localized macroscopic objects in general. Instead of particles, there hence is a field of density of stuff (gunk), with there being more stuff at some points and regions of space than at others.

Given that there is just one universal wave-function, there is just one matter density field in the whole of space. As in Bohmian mechanics, one can take the universal wave-function as it figures in the GRW law to represent a non-local beable over and above the local beables that consist in the density of stuff at points of space. Again, one can spell out this notion of a non-local beable in terms of a causal property or a disposition that determines the temporal development of the matter density field in physical space, the manner in which the matter density develops in time being the effect of that causal property or the manifestation of that disposition. Since the GRW law is probabilistic instead of deterministic, that disposition is a propensity grounding objective probabilities for a certain temporal development of the density of matter in the whole of space (see Dorato and Esfeld 2010 for dispositions and propensities in GRW quantum mechanics). In any case, again, there is a non-local common cause of the temporal development of the density of matter in the whole of space.

The main ontological difference between Bohmian mechanics and the GRW mass density theory is not that the law of motion of the latter is probabilistic, whereas the law of the former is deterministic, but that according to the latter theory, physical entities do not always move continuously through space. Consider, for the sake of illustration, Einstein's thought experiment with one particle in a box: the box is split in two halves which are sent in opposite directions, say from Brussels to New York and Tokyo. When the half-box arriving in New York is opened and found to be empty, all accounts of quantum mechanics that recognize the uniqueness of measurement outcomes agree that the particle is in the half-box in Tokyo (see Norsen 2005 for details of this thought experiment). On Bohmian mechanics, the particle always travels in one of the two half-boxes, depending on its initial position, and the cause of its motion is local in this case, given by the effective wave-function of the particle. On the GRW mass density theory, the particle is in fact a field that stretches over the whole box and that is split in two halves of equal density when the box is split, these masses travelling in opposite directions. Upon interaction with a measurement device, one of these masses (the one in New York in the example given above) disappears, while the mass density in the other half-box (the one in Tokyo) increases so that the whole mass is concentrated in one of the half-boxes. The difference between Bohmian mechanics and the GRW mass density theory thus is the following one: on the latter theory, non-local causation means that a physical entity disappears in one place and appears in another; on the former theory, non-local causation means that a causal property, which is non-local in that it is instantiated by a configuration of particles as a whole, determines the velocity of each particle, while particles always move through space on continuous trajectories.

One might be tempted to describe what happens according to the GRW mass density theory in the above example by saying that some matter travels from New York to Tokyo when one of the half-boxes is opened, but the considerations of section 2 speak against assigning any finite speed to this travel. Consequently (recall van Fraassen's remark about "instantaneous travel" quoted in section 2), use of the "travel" metaphor is inappropriate. For lack of a better term, let us say that some matter is *delocated* from New York to Tokyo. Nevertheless, what is known as the collapse of the wave-function does not have to be instantaneous, but can be a continuous process in the sense that the temporal development of the wave-function does not display any "jumps". The most important proposal in this respect is the *continuous spontaneous localization model* (see Ghirardi, Pearle and Rimini 1990). However, this model does not postulate a finite speed of the causal influence either; the time it takes for the matter density to disappear in one place and to appear in another place does not depend on the distance between the two places. Therefore, there is no travel of matter upon wave-function collapse on this model either. It hence does not face the objections listed in section 2. Another advantage of the continuous spontaneous localization model is that the wave-function can, as in theories without collapse, be described by a differential equation conforming to the general scheme (2). This makes it possible to conceive the temporal development of the mass density as a causal process in the sense of Blondeau and Ghins (2012).

Let us now turn to the explanation that the GRW mass density theory gives of the EPR experiment. Recall that as in Bohmian mechanics, there is no local beable of spin in this theory: the only local physical property (local beable) is the density of mass at any given point of space. Thus, the outcomes of what is regarded as measurements of spin have to be accounted for in terms of the temporal development of the mass density. Furthermore, since the GRW equation is probabilistic instead of deterministic, a measurement outcome is not fixed by anything before the wave-function has actually undergone a spontaneous localization in configuration space.

The interaction of the two-"particle" matter field with both the apparatuses triggers an irreducibly stochastic process (the "collapse") changing the shape of the mass density, which then causes the two measurement outcomes. Once again, the common cause is non-local, as it depends on the interaction of the matter field with both of the measurement devices and the non-local beable that is represented by the wave-function. Hence, although the GRW theory violates the condition known as outcome independence (and not, like Bohmian mechanics, the condition known as parameter independence, as explained in the previous section), this does not mean that one of the outcomes causes the other one. Instead, there is a common cause given by the mass density of the two-"particle" system, namely its delocation as represented by the spontaneous localization of the wave-function in configuration space (the "collapse").

Although the GRW theory is a paradigm case of an indeterministic theory, it is important to note that the common cause in the explanation just given acts deterministically: once the "collapse" occurs, the propensity for spontaneous localization having been triggered, the shape of the matter field completely determines the measurement outcomes. If, alternatively, one were to identify the state of the system *before* the spontaneous localization of the wave-function as a (probabilistic) cause of the outcomes, it would not be legitimate to refer to it as a common cause, because the pre-collapse state in general fails to screen off the EPR correlations. Therefore, it is the delocation of the mass density in "collapse" that acts as a (deterministic, non-local) common cause of the measurement outcomes according to the GRW mass density theory.

As in the case of Bohmian mechanics, the non-locality involved in this theory does not allow for the sending of superluminal signals. But unlike Bohmian mechanics, the GRW mass density theory does not need to invoke fine-tuning to explain why this is so. The reason is that the GRW theory does not postulate additional variables (such as Bohmian particle positions) that then need to be coordinated with the statistical role of the wave-function via an additional postulate (i.e., the quantum equilibrium hypothesis). The mass density is completely specified by the wave-function, and the impossibility of superluminal signalling is accounted for by the fact that the pre-collapse state of the mass density cannot be observed, since any intervention by means of a macroscopic device brings about an irreducibly stochastic change of the mass density distribution.

Let us briefly mention the fact that there is another ontology of the GRW formalism available, which gives up the idea of a continuous distribution of matter in space and of a physical entity being delocated across space. According to the ontology that Bell (2004, ch. 22) proposed for the GRW formalism, whenever there is a spontaneous localization of the wave-function in configuration space, that development of the wave-function in configuration space represents an event occurring in physical space, namely there being a flash centred around a space-time point (the term "flash", however, is not Bell's, but was introduced by Tumulka 2006). The flashes are all there is in space-time. That is to say, apart from when it spontaneously localizes, the temporal development of the wave-function in configuration space does not represent the distribution of matter in physical space. It represents the objective probabilities for the occurrence of further flashes, given an initial configuration of flashes and an initial wave-function. Hence, in contrast to Bohmian mechanics and the GRW mass density ontology, the GRW flash ontology does not admit a continuous distribution of matter: there are only flashes being sparsely distributed in space-time, but no trajectories (worldlines) or fields (densities) of anything in physical space. Nonetheless, one may seek a causal explanation of the correlated outcomes of the EPR experiment also in the framework of the GRW flash ontology. In this case, however, one would have to conceive the non-local common cause as stretching back into the past, including notably the flashes at the source of the experiment.

In conclusion, calling for a causal explanation of experimental results is a well-taken demand also in the case of the correlated outcomes of EPR-type experiments, and that demand can be satisfied. In particular, the two main versions of quantum physics that recognize the existence of measurement outcomes – Bohmian mechanics and the GRW mass density theory – each offer an explanation of the results of EPR-type experiments in terms of a non-local common cause.

**Acknowledgments** Parts of this paper were presented at the *PhiloSTEM-5* Workshop (Fort Wayne IN, March 2013) and the Conference on *Causality and Experimentation in the Sciences* (Paris, July 2013). We wish to thank the participants of these events, in particular Charles Sebens and Phil Dowe, for insightful comments. We are also grateful to Sheldon Goldstein and two anonymous referees for helpful remarks.